\documentclass[aps,prd,preprint,
superscriptaddress,preprintnumbers,
showpacs,nofootinbib,nobibnotes]{revtex4}
\usepackage{amsfonts,amssymb,amsmath}
\usepackage{graphicx}

\newcommand{\be}{\begin{equation}}
\newcommand{\bea}{\begin{eqnarray}}
\newcommand{\ee}{\end{equation}}
\newcommand{\eea}{\end{eqnarray}}
\newcommand{\bpi}{\begin{picture}}
\newcommand{\bce}{\begin{center}}
\newcommand{\epi}{\end{picture}}
\newcommand{\ece}{\end{center}}
\newcommand{\D}{\displaystyle}
\def\chic#1{{\scriptscriptstyle #1}}

\def\gb{{\rm I}\hspace{-0.07cm}\Gamma}

\newcommand{\de}{{\Delta}_{\rm \mbox{\tiny E}}}  
\newcommand{\da}{\rm \mbox{\tiny E}}  

\newcommand{\Valencia}{Departamento de F\'\i sica Te\'orica
and IFIC, Centro Mixto,\\ Universidad de Valencia (Fundaci\'on General) -- CSIC \\
E-46100, Burjassot, Valencia, Spain}

\begin{document}


\title{Infrared finite ghost propagator in the Feynman gauge}

\author{A.~C.~Aguilar}
\affiliation{\Valencia}

\author{J.~Papavassiliou}
\affiliation{\Valencia}

\begin{abstract} 

We demonstrate how to obtain from the Schwinger-Dyson equations of QCD
an infrared  finite ghost  propagator in the  Feynman gauge.   The key
ingredient in this construction is the longitudinal form factor of the
non-perturbative gluon-ghost  vertex, which, contrary  to what happens
in the Landau gauge, contributes  non-trivially to the gap equation of
the ghost.   The detailed study  of the corresponding  vertex equation
reveals that in the presence  of a dynamical infrared cutoff this form
factor remains finite in the limit of vanishing ghost momentum.  This,
in turn, allows  the ghost self-energy to reach a  finite value in the
infrared, without  having to assume any additional  properties for the
gluon-ghost  vertex, such  as  the presence  of  massless poles.   The
implications of this result and possible future directions are briefly
outlined.

\end{abstract}


\maketitle

\setcounter{section}{0}
\section{Introduction}
\label{Sect:Intro}

The non-perturbative properties of  the basic Green's functions of QCD
have been the focal point  of intensive scrutiny in recent years, with
particular emphasis  on the propagators of the  fundamental degrees of
freedom,  gluons, quarks, and  ghosts.  Even  though it  is well-known
that  these quantities  are not  physical,  since they  depend on  the
gauge-fixing scheme and parameters used  to quantize the theory, it is
generally accepted that reliable information on their non-perturbative
structure is essential for unraveling the infrared (IR) dynamics of QCD.

There are two main tools usually employed in this search: the lattice,
where  space-time is discretized  and the  quantities of  interest are
evaluated   numerically \cite{Creutz:1980zw, Mandula:1987rh, Bernard:1993tz},   and   the 
 intrinsically   non-perturbative
equations governing  the dynamics of  the Green's functions,  known as
Schwinger-Dyson equations (SDE) 
\cite{Dyson:1949ha, Schwinger:1951ex, Cornwall:1974vz, Marciano:1977su}.  
In principle,  the lattice includes
all non-perturbative  features and  no approximations are  employed at
the level  of the  theory.  In practice,  the main  limitations appear
when  attempting  to  extrapolate  the results  obtained  with  finite
lattice volume to the continuous space-time limit.  On the other hand,
the main difficulty with the SDE  has to do with the need to devise  
a self-consistent truncation  scheme that preserves crucial
field-theoretic properties,  such as  the transversality of  the gluon
self-energy,   known    to   be   valid    both   perturbatively   and
non-perturbatively, as a consequence of the BRST symmetry~\cite{Becchi:1975nq}.

Significant  progress has  been accomplished on this last issue
due to the development of the  truncation scheme that is based on the
all-order   correspondence~\cite{Binosi:2002ft}   between  the   pinch
technique (PT)~\cite{Cornwall:1981zr, Cornwall:1989gv}
and the Feynman  gauge of the Background  Field Method
(BFM)~\cite{Abbott:1980hw}.  One of its most powerful features 
is  the special  way in which  the transversality of  the gluon
self-energy is realized.  Specifically,  by virtue of the Abelian-like
WIs   satisfied  by   the   vertices  involved,   gluonic  and   ghost
contributions are {\it separately} transverse, within {\it each} order
in the ``dressed-loop'' expansion of the SDE~\cite{Aguilar:2006gr} for
the gluon propagator.  This property, in turn, allows for a systematic
truncation  of  the  full  SDE, preserving  
at every step the
crucial property of gauge invariance.

The first  approximation to the  SDE of the gluon  propagator involves
the one-loop  dressed gluonic  graphs only, since  in this  scheme the
ghost loops may be  omitted without compromising the transversality of
the  answer.   As is  well-known,  the Feynman  gauge  of  the BFM  is
particularly privileged,  being dynamically  singled out as  the gauge
that  directly encompasses  the  relevant gauge  cancellations of  the
PT~\cite{Binosi:2002ft}.    Therefore,  the   aforementioned  one-loop
dressed  graphs have been  considered in  this particular  gauge.  The
detailed  study  of the  resulting  integral  equation  for the  gluon
propagator gave rise  to solutions that reach a  {\it finite} value in
the   deep    IR~\cite{Aguilar:2006gr,   Aguilar:2007ie}.    Following
Cornwall's   original  idea~\cite{Cornwall:1979hz,Cornwall:1981zr}  of
describing  the  IR sector  of  QCD in  terms  of  an effective  gluon
mass~\cite{Bernard:1981pg,  ref},  these  solutions have  been  fitted
using ``massive''  propagators of the  form $\Delta^{-1}(q^2) =  q^2 +
m^2(q^2)$,  with  $m^2(0)>0$,  and  the  crucial  characteristic  that
$m^2(q^2)$ is not ``hard'',  but depends non-trivially on the momentum
transfer  $q^2$.    In  addition,  finite  solutions   for  the  gluon
propagator in the  Landau gauge have been reported  in various lattice
studies~\cite{Alexandrou:2000ja},  and were  recently  confirmed using
lattices  with  significantly larger  volumes~\cite{Cucchieri:2007md}.


Even though  the omission of  the ghost loops within  this formulation
does not introduce any artifacts,  such as the loss of transversality,
the actual  behavior of the  ghosts may change the  initial prediction
for   the  gluon   propagator,  not   just  quantitatively   but  also
qualitatively.  For  example, an IR  divergent solution for  the ghost
propagator could destabilize the  finite solutions found for the gluon
propagator.   Therefore,   a  detailed   study  of  the   ghost  sector
constitutes the next  challenge in this approach. In  the present work
we will  consider the SDE  for the ghost  sector in the  (BFM) Feynman
gauge, in order to  complement the corresponding analysis presented in
~\cite{Aguilar:2006gr,  Aguilar:2007ie} in  the same  gauge.   The BFM
Feynman  rules are  in general  different  to those  of the  covariant
renormalizable   gauges~\cite{Abbott:1980hw};  in   the   former,  for
example, in addition to the bare gluon propagator, the bare three- and
four-gluon vertices involving background  and quantum gluons depend on
the (quantum)  gauge fixing  parameter.  Notice, however,  that, since
there are  no background  ghosts, the Feynman  rules relevant  for the
ghost  sector are  identical  to  both the  covariant  gauges and  the
BFM. Therefore, the analysis and the results presented in this article
carries over directly to the conventional Feynman gauge.

In  this article  we  demonstrate  that the  ghost  propagator in  the
Feynman   gauge  can   be  made   finite  in   the  IR,   through  the
self-consistent treatment of the  gluon-ghost vertex and the ghost gap
equations.   The  key  ingredient  that  makes this  possible  is  the
``longitudinal''  form-factor in  the tensorial  decomposition  of the
gluon-ghost vertex,  $\gb_{\mu}^{bcd}(p,q,k)$, i.e.  the  co-factor of
$k_{\mu}$, where $k$ is the four-momentum of the gluon; evidently this
term  gets  annihilated  when  contracted with  the  usual  transverse
projection  operator.  As we  will explain  in detail,  this component
acquires a special role for  all values of the gauge fixing parameter,
with  the very  characteristic  exception of  the  Landau gauge.   The
reason is simply that in  the Landau gauge the entire gluon propagator
is transverse, both its self-energy and its free part, whereas for any
other  value  of the  gauge-fixing  parameter  the  free part  is  not
transverse.  As a result, when the gluon-ghost vertex is inserted into
the SDE for  the ghost propagator, $D(p^2)$, its  part proportional to
$k_{\mu}$ dies when contracted with the gluon propagator in the Landau
gauge; however, in any other gauge it survives due to the free-part of
the gluon  propagator.  The resulting contribution  has the additional
crucial  property of  not vanishing  as the  external momentum  of the
ghost goes to zero.  Therefore, contrary to what happens in the Landau
gauge  where only  the part  of the  vertex proportional  to $p_{\mu}$
survives, one does  {\it not} need to assume  the presence of massless
pole terms of the form $1/p^2$ in order to obtain a nonvanishing value
for  $D^{-1}(0)$.    Instead,  the   only  requirement  is   that  the
longitudinal form factor simply does not vanish in that limit.

The paper is organized as follows: In section~\ref{Sect:pghost} we set
up the SDE for the ghost propagator, assuming the most general Lorentz
structure    for     the    fully    dressed     gluon-ghost    vertex
$\gb_{\mu}^{bcd}(p,q,k)$.  We then  discuss under  what  condition the
resulting  expression may yield  a finite  value for  $D^{-1}(0)$, and
analyze  the   profound  differences   between  the  Landau   and  the
Feynman-type of gauges.   In section~\ref{Sect:vertex} we first derive
the gluon-ghost vertex  under certain simplifying assumptions, discuss
in   detail   the  approximations   employed.   Next   we  study   its
non-perturbative solutions employing  various physically motivated, IR
finite  Ans\"atze  IR  for   the  gluon  and  ghost  propagators.   In
Section~\ref{Sect:comb}  we combine  the results  of the  previous two
sections,  deriving the self-consistency  condition necessary  for the
system  of  equations to  be  simultaneously  satisfied.  Finally,  in
section~\ref{Sect:Concl}  we  discuss  our  results  and  present  our
conclusions.



\setcounter{equation}{0}
\section{General considerations on the IR behavior of the ghost}
\label{Sect:pghost}

In this section we derive the SDE for the ghost propagator $D(p^2)$ in
a general covariant gauge, and study qualitatively its predictions for
 $D(0)$ for  various gauge choices.   In particular, we  establish that
away from the  Landau gauge the ghost propagator  may acquire a finite
value at the  origin, without the need to  assume a singular IR behavior
for the form factors of  the fully dressed ghost-gluon vertex entering
into  the SDE.  Our attention  will  eventually focus  on the  Feynman
gauge, which, as mentioned in  the Introduction, is singled out within
the PT-BFM scheme.

\begin{figure}[ht]
\includegraphics[scale=0.7]{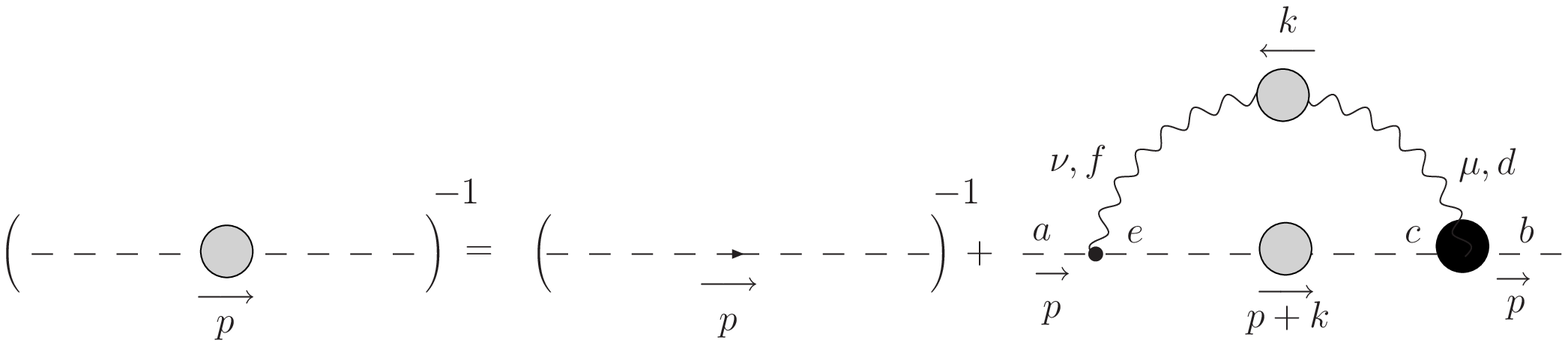}
\caption{ The SDE of the ghost propagator.}
\label{SDE_GHOST}
\end{figure}

The full ghost propagator $D^{ab}(p)$ is usually written in the form
\be
D^{ab}(p)= i\delta^{ab}D(p)\,,
\label{fghost}
\ee
and the SDE satisfied by $D(p^2)$,  depicted diagrammatically in Fig.\ref{SDE_GHOST}, reads 
\be
D^{-1}(p^2)= p^2 +iC_{\rm A}g^2 \int\![dk]\,\Gamma^{\mu}\Delta_{\mu\nu}(k)\gb^{\nu}(p,p+k,k)D(p+k) \,.
\label{lgluon}
\ee
We have used $f^{acd}f^{bcd}=\delta^{ab}C_{\rm A}$, with $C_{\rm
A}$  the Casimir  eigenvalue  in the  adjoint representation  [$C_{\rm
A}=N$  for $SU(N)$], and have introduced  the short-hand
notation  $ [dk]  = d^d  k/(2\pi)^d$ \,,  where $d=4-\epsilon$  is the
dimension of space-time used in dimensional regularization. 
$\Delta_{\mu\nu}(k)$ is the fully dressed gluon propagator, whereas
$\gb$ denotes the fully dressed gluon-ghost vertex and 
$\Gamma$ its tree-level value.

Specifically, in the covariant gauges the  full  gluon  propagator
$\Delta_{\mu\nu}^{df}(k) =-i\delta^{df}\Delta_{\mu\nu}(k) $
has the general form
\begin{equation}
\Delta_{\mu\nu}(k)= \left[{\rm P}_{\mu\nu}(k)\Delta(k^2) + 
\xi\frac{k_{\mu}k_{\nu}}{k^4}\right]\,,
\label{prop_cov}
\end{equation}
where
\be
{\rm P}_{\mu\nu}(k)= \ g_{\mu\nu} - \frac{\D k_\mu
k_\nu}{\D k^2}\,,
\label{projector}
\ee
is the transverse projector, and $\xi$ is the gauge fixing parameter;
$\xi=1$ corresponds to the Feynman gauge and $\xi=0$
to the Landau gauge.
The scalar function $\Delta(k^2)$ is related to the 
all-order gluon self-energy $\Pi_{\mu\nu}(k)$,
\be
\Pi_{\mu\nu}(k)={\rm P}_{\mu\nu}(k) \Pi(k^2)\,,
\ee
through
\be
\Delta^{-1}(k^2) = k^2 + i \Pi(k^2)\,.
\label{fprog}
\ee

The bare gluon-ghost vertex appearing in (\ref{lgluon})
is given by $\Gamma_{\mu}^{eaf}=-gf^{eaf}q_{\mu}$, with (\mbox{$q=p+k$}).
Choosing $p_{\mu}$ and $k_{\mu}$ as the two linearly independent four-vectors, 
the most general decomposition 
for the fully dressed gluon-ghost vertex $\gb_{\nu}^{bcd}(p,q,k)$ is expressed as~\cite{Pascual:1980yu}
\bea
\gb_{\mu}^{bcd}(p,q,k)&=&-gf^{bcd}\,\gb_{\mu}(p,q,k) \,, \nonumber  \\
\gb_{\mu}(p,q,k)&=& A(p^2,q^2,k^2)p_{\mu} + B(p^2,q^2,k^2)k_{\mu} \,,
\label{fvg}
\eea
where $k$ is the  outgoing  gluon momentum,  and  $p$, $q$  the
outgoing and  incoming ghost momenta,  respectively.  The dimensionless
scalar  functions $A(p^2,q^2,k^2)$ and  $B(p^2,q^2,k^2)$ are  the form
factors  of the  gluon-ghost vertex.  In particular,  notice  that the
tree-level result  is recovered when  we set \mbox{$A(p^2,q^2,k^2)=1$}
and \mbox{$B(p^2,q^2,k^2)=0$}.
Finally, it is important to emphasize that all fully-dressed scalar quantities 
($D$, $\Delta$, $A$, and $B$) 
depend explicitly (and non-trivially) on  
the value of the gauge-fixing parameter $\xi$ already at the 
level of one-loop perturbation theory. 

It  is  then  straightforward   to  derive  the  Euclidean  version  of
Eq.(\ref{lgluon}); to that end, we  set $p^2= -p^2_{\da}$,  define $
\de(p^2_{\da})= -  {\Delta}(-p^2_{\da})$, and $  D_{\da}(p^2_{\da})= -
D(-p^2_{\da})$  , and  for the  integration  measure we  have $[dk]  =
i[dk]_{\chic  E}= i  d^4  k_{\chic E}/(2\pi)^4  $.  
Suppressing the subscript ``E'' everywhere 
except in the integration measure, and without  any
assumptions  on the   functional  form   of
\mbox{$A(p^2,q^2,k^2)$}  and \mbox{$B(p^2,q^2,k^2)$}, 
the ghost SDE of Eq.~(\ref{lgluon}) becomes

\bea
D^{-1}(p^2)&=&p^2 -C_{\rm A}g^2\!\!\int[dk]_{\da}\,\left[p^2-\frac{(p \cdot k)^2}{k^2}
\right]A(p^2,q^2,k^2)\,\Delta(k)\,D(p+k) 
\nonumber \\
&&
-C_{\rm A}g^2 \xi\!\!\int[dk]_{\da}\;\frac{p\cdot k}{k^2}\left[A(p^2,q^2,k^2)+B(p^2,q^2,k^2)+
\frac{p\cdot k}{k^2}\,\right]\,D(p+k) 
\nonumber \\
&&
- C_{\rm A}g^2\xi\!\!\int[dk]_{\da}\,B(p^2,q^2,k^2)\,D(p+k) 
 \,,
\label{g1}
\eea

As a check, we can recover from (\ref{g1})
the one-loop result for the ghost propagator in the Feynman gauge ($\xi=1$) by 
substituting the tree-level expressions for the ghost and gluon propagators 
and setting  \mbox{$A(p^2,q^2,k^2)=1$} and \mbox{$B(p^2,q^2,k^2)=0$}; specifically
\be
D^{-1}(p^2)=p^2\left[1+\frac{C_{\rm A}g^2}{32\pi^2}\ln\left(\frac{p^2}{\mu^2}\right)\right] \,.
\label{1loop}
\ee

In order  to obtain from  (\ref{g1}) the behavior of  $D(p^2)$ for
the full  range of the momentum  $p^2$ one needs  to provide additional
information    for   the    forms    factors   $A(p^2,q^2,k^2)$    and
$B(p^2,q^2,k^2)$, obtained from the corresponding SDE satisfied by the
gluon-ghost vertex. Thus, the  complete treatment of this problem would
require the solution of a complicated system of coupled SDE.  However,
several interesting conclusions about the IR behavior of $D(p^2)$ may
be  drawn,  by considering  the  qualitative  behavior  of the  forms
factors $A(p^2,q^2,k^2)$ and $B(p^2,q^2,k^2)$, as $p\to 0$.

We start  by considering what happens  in the Landau  gauge.  First of
all, let us  assume that the various quantities  appearing on
the r.h.s. of (\ref{g1}) are regular functions of $\xi$~\cite{foot1}.
Then,  if we set  $\xi=0$, only  the first  integral on the  
r.h.s.  of  (\ref{g1}) survives; thus, $D^{-1}(p^2)$  is only affected
by  the  functional  form  of  $A(p^2,q^2,k^2)$.  In  particular,  the
behavior  of  $D(p^2)$ as  $p\rightarrow  0$  will  depend on  whether
$A(p^2,q^2,k^2)$ is divergent or finite  in that limit, i.e. on whether
or  not  $A(p^2,q^2,k^2)$  contains  $(1/p^2)$ terms.   Evidently,  if
$A(p^2,q^2,k^2)$    does   not   contain    poles,   one    has   that
${\displaystyle\lim_{p\to  0}}\,D^{-1}(0)=0$, and therefore  the ghost
propagator  will  be  divergent in  the  IR.  On  the other  hand,  if
$A(p^2,q^2,k^2)$ contains $(1/p^2)$ terms, ${\displaystyle\lim_{p\to
0}}\,D^{-1}(0)\neq  0$  allowing  for  finite solutions  for  the  ghost
propagator.

According  to this  general  argument,  the only  way  for getting  an
IR-finite  propagator  in  the   Landau  gauge  is  by  assuming  that
$A(p^2,q^2,k^2)$ contains poles \cite{Fischer:2006ub, Boucaud:2007va}.   
However, lattice simulations in the
Landau   gauge   seem   to   favor  a   IR-finite   $A(p^2,q^2,k^2)$;
specifically, it  was found that deviations of  the gluon-ghost vertex
from  its   tree-level  value   are  very  small   in  the   IR,  i.e.
$A(p^2,q^2,k^2)\approx 1$ \cite{Cucchieri:2004sq}.  In addition, a  detailed study of  the SDE
equation for  $\gb$ in the same  gauge shows no  singular behavior for
$A(p^2,q^2,k^2)$ \cite{Schleifenbaum:2004id} . 
These  findings appear  to be consistent  with recent
lattice results on the non-perturbative structure of the ghost propagator,
which indicate that  $D^{-1}(p^2)$ in the Landau gauge  diverges, at a
rate that deviates only mildly from the tree-level
 expectation of $1/p^2$~\cite{Cucchieri:2007md}.

Evidently, the picture for $\xi \neq 0$  
is drastically different. Indeed, away from the Landau gauge
the r.h.s of  (\ref{g1}) 
involves both  form
factors, $A(p^2,q^2,k^2)$ and  $B(p^2,q^2,k^2)$. 
Moreover, unlike the first two terms, the third one does not 
contain any kinematic factors proportional to $p$. Thus, 
in order for it not to vanish as $p\to 0$ one does not need to assume
any singular structure for $B(p^2,q^2,k^2)$; instead, 
it is sufficient to simply have that $B(0,k^2,k^2) \neq 0$.

After this key observation, 
we will take the limit of 
of Eq.~(\ref{g1}) as $p\to 0$, assuming that  
$A(p^2,q^2,k^2)$ does not contain $(1/p^2)$ terms.
Focusing for 
concreteness on the physically relevant case of $\xi=1$, we find that 
 in the aforementioned kinematic limit Eq.(\ref{g1}) reduces to 
\be
D^{-1}(0)=-C_{\rm A}g^2\!\!\int[dk]_{\da}\,B(0,k^2,k^2)\,D(k) \,. 
\label{tadghost}
\ee
Of course, if the assumption that 
$A(p^2,q^2,k^2)$ is regular as $p\to 0$ does not hold, then
the other integrals will also contribute to the r.h.s. of (\ref{tadghost}).
However, modulo the rather contrived scenario of  
fine-tuned cancellations, the r.h.s. will still be different from zero.
Evidently, from (\ref{tadghost}) we deduce that 
if $B(0,k^2,k^2)=0$ than $D^{-1}(0)=0$.
On the other hand, if $B(0,k^2,k^2) \neq0$, i.e.
if it does not vanish identically, 
then one may have 
a non-vanishing $D^{-1}(0)$.
Of course, having a non-vanishing $B(0,k^2,k^2)$
is not a sufficient condition for $D^{-1}(0)\neq 0$;
one has to assume in addition that (i) the integral on the r.h.s. of (\ref{tadghost}).
is convergent, or it can be made convergent through 
proper regularization, and (ii) that the 
integral is not zero due to some other, rather contrived circumstances
(for instance, if $B(0,k^2,k^2)$ turned 
out not to be a monotonic function, the various contributions 
from different integration regions could cancel against each other).

An explicit calculation may confirm that 
\mbox{$B(0,k^2,k^2)$} vanishes at one-loop \cite{Watson:1999ha}, and it is reasonable to expect 
this to  persist to all orders in perturbation theory. 
 Therefore, in what follows we will examine the possibility that  
$B(0,k^2,k^2)$  may not vanish  non-perturbatively.
In particular, we will study the SDE
determining \mbox{$B(p^2,q^2,k^2)$} for the special  kinematic configuration 
appearing in (\ref{tadghost}), namely
where the outgoing ghost momentum, $p$, is set equal to zero 
(i.e. $p=0$ and $q=k$). 
In the context of the linearized 
approximation that we employ in the next section 
this kinematic configuration offers the particular technical advantage
of dealing with a function of only one variable instead of two.

\setcounter{equation}{0}
\section{The gluon-ghost vertex}
\label{Sect:vertex}

\begin{figure}[ht]
\includegraphics[scale=0.7]{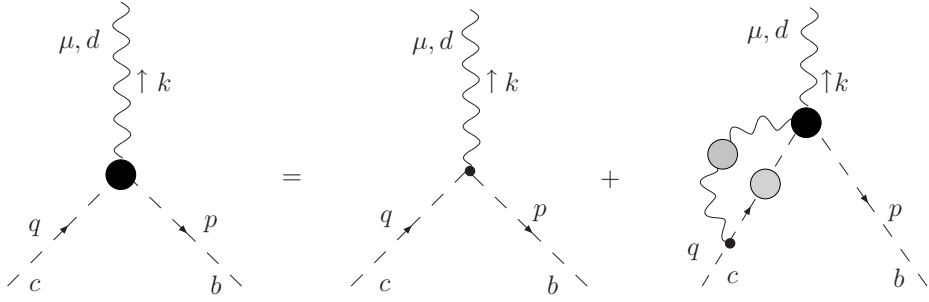}
\caption{SDE for the gluon-ghost vertex.}
\label{SDE1}
\end{figure}

In this section we set up and solve, after certain 
simplifying approximations, the SDE governing the 
behavior of the form  factor \mbox{$B(0,k^2,k^2)$}. 
This can be done  by taking the following limit  of the gluon-ghost vertex,
$\gb_{\mu}(p,q,k)$ ,
\be
B(0,k^2,k^2) = \lim_{p\rightarrow 0} \left[ \frac{1}{k^2}k^{\mu}\,\gb_{\mu}(p,q,k)\right]\,.
\label{b0}
\ee
where $\gb_{\mu}(p,q,k)$ obeys the SDE \cite{Marciano:1977su} represented in Fig.{\ref{SDE1}}

We next introduce some approximations regarding the 
form of the two-ghost--two-gluon scattering kernel, appearing on the r.h.s. of Fig.(\ref{SDE1}).
The first approximation
is to keep only the lowest order contributions in its skeleton expansion, i.e. we 
expand the aforementioned kernel in terms of the 1PI fully dressed three-particle 
vertices of the theory, neglecting 
diagrams that contain four-point functions. 

\begin{figure}[hb]
\includegraphics[scale=0.7]{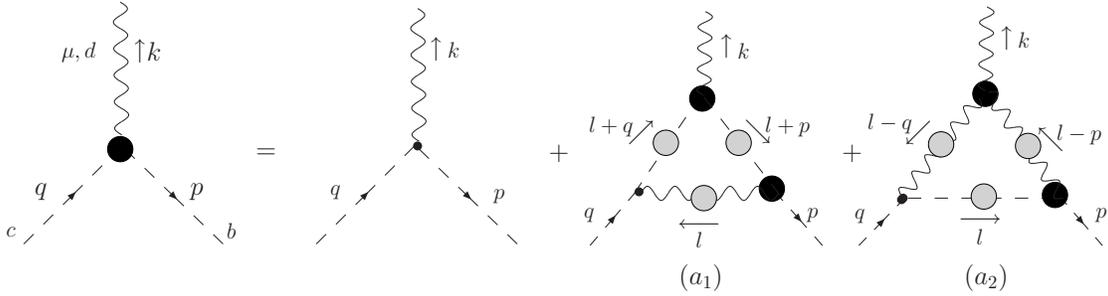}
\caption{Truncated version of the SDE for the gluon-ghost vertex.}
\label{SDE}
\end{figure}

We then arrive at the truncated SDE shown in  Fig.(\ref{SDE}), which reads,
%
%
%
%
\be
\gb_{\mu}^{bcd}(p,q,k)=\Gamma_{\mu}^{bcd} + \gb_{\mu}^{bcd}(p,q,k)\left.\right|_{a_1} 
+\gb_{\mu}^{bcd}(p,q,k)\left.\right|_{a_2} \,,
\label{fullg} 
\ee
where the closed expressions corresponding to the diagrams ($a_1$) and ($a_2$) are given by
\bea
\gb_{\mu}^{bcd}\left.\right|_{a_1}&=&\int\![dl]\,\gb_{\mu}^{emd}(l+p,l+q,k)D_{ee^{\prime}}(l+p)
\,\gb_{\nu^{\prime}}^{be^{\prime}n^{\prime}}(p,l+p,l)\,\Delta^{\nu\nu^{\prime}}_{nn^{\prime}}(l)\,
\Gamma_{\nu}^{m^{\prime}cn}\,D_{mm^{\prime}}(l+q)  \,,\nonumber \\
\gb_{\mu}^{bcd}\left.\right|_{a_2}&=&\int\![dl]\,\gb_{\mu\nu\sigma}^{dem}(-k,q-l,l-p)
\Delta^{\sigma\sigma^{\prime}}_{mm^{\prime}}(l-p)\,
\gb_{\sigma^{\prime}}^{bn^{\prime}m^{\prime}}(p,l,l-p)\,D_{nn^{\prime}}(l)\,
\Gamma_{\nu^{\prime}}^{nce^{\prime}}\Delta_{ee^{\prime}}^{\nu\nu^{\prime}}(l-q) \,,\nonumber \\
\label{contr}
\eea
with the momentum routing as given in Fig.(\ref{SDE}).

\begin{figure}[ht]
\includegraphics[scale=0.7]{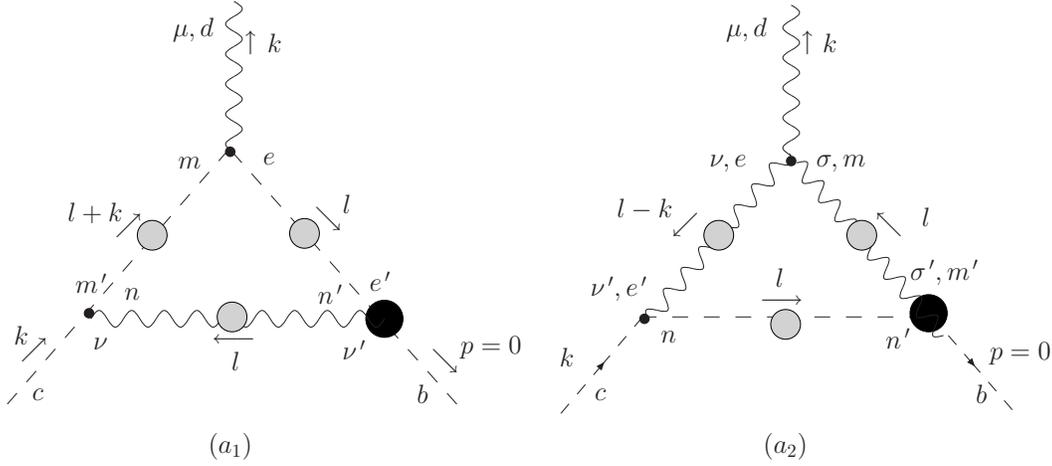}
\caption{Contributions for the gluon-ghost vertex equation in the limit of $p\rightarrow 0$.}
\label{vertex}
\end{figure}

Our next approximation is to linearize the equation by  
substituting in (\ref{contr}) 
\mbox{$\gb_{\mu}^{emd}(l+p,l+q,k)$} and $\gb_{\mu\nu\sigma}^{dem}(-k,q-l,l-p)$
by their bare, tree-level expressions. Since we are eventually interested 
in the limit of the equation as $p\to 0$, this amounts finally to 
the replacement 
\bea
&\gb_{\mu}^{emd}(l+p,l+q,k)&\rightarrow -gf^{emd}l_{\mu} \,, \nonumber  \\
&\gb_{\mu\nu\sigma}^{dem}(-k,q-l,l-p)& \rightarrow gf^{dem}[(2l-k)_{\mu}g_{\nu\sigma} -(k+l)_{\nu}g_{\mu\sigma}
+(2k-l)_{\sigma}g_{\mu\nu}]\,. 
\label{two_vertices}
\eea
in diagrams $(a_1)$ and $(a_2)$, respectively. 
The diagrammatic representation of the resulting contributions at $p\rightarrow 0$ is given in
Fig.(\ref{vertex}).

Factoring out the color structure by using the standard identity
\mbox{$f^{axm}f^{bmn}f^{cnx}  = \frac{1}{2}\, C_A f^{abc}$},
it is easy to verify that in the limit $p\rightarrow 0$ the linearized 
version of Eq.(\ref{contr}) reads
\bea
\gb_{\mu}^{bcd}(0,k,k)\left.\right|_{a1}&=& if^{bcd}\frac{C_{\rm A}g^3}{2} \int\![dl]
\,l_{\mu}(l+k)_{\nu^{\prime}}l_{\nu}\Delta^{\nu\nu^{\prime}}(l)B(0,l^2,l^2)D(l)D(l+k) \,,\nonumber \\
\gb_{\mu}^{bcd}(0,k,k)\left.\right|_{a2}&=&
-if^{bcd}\frac{C_{\rm A}g^3}{2} \int\![dl]\,\Gamma_{\mu\nu\sigma}\,l_{\nu^{\prime}} l_{\sigma^{\prime}}
\Delta^{\sigma\sigma^{\prime}}(l)\Delta^{\nu\nu^{\prime}}(l-k)B(0,l^2,l^2)D(l) \,.
\label{contr2}
\eea

Since the bare gluon-ghost is proportional to $p_{\mu}$, it follows immediately
from Eqs.(\ref{b0}), (\ref{fullg}) and (\ref{contr2}), that 
\bea
B(0,k^2,k^2)&=&\frac{k^{\mu}}{k^2}\Big[\gb_{\mu}(0,k,k)\left.\right|_{a_1}
+\gb_{\mu}(0,k,k)\left.\right|_{a_2}\Big]\,,\nonumber \\
k^{\mu}\gb_{\mu}(0,k,k)\left.\right|_{a_1}&=& -\frac{i}{2}C_{\rm A}g^2 
\int\![dl]\,\left[k\cdot l+\frac{(k\cdot l)^2}{l^2}\right]B(0,l^2,l^2)D(l)D(l+k)
\,,\nonumber \\
k^{\mu}\gb_{\mu}(0,k,k)\left.\right|_{a_2}&=& 
+\frac{i}{2}C_{\rm A}g^2 
\int\![dl]\left[\frac{(k\cdot l)^2}{l^2}-k^2\right]B(0,l^2,l^2)D(l)\Delta(l+k)
\,.
\label{contr3}
\eea
The Euclidean version of (\ref{contr3}) can be  easily derived using the same rules as before, 
leading to
\bea
B(0,k^2,k^2)&=&-\frac{C_{\rm A}g^2}{32\pi^4}\left\{ \frac{1}{k^2}\int\!\!d^{4} l\,\frac{(k\cdot l)^2}{l^2}
B(0,l^2,l^2)D(l)\big[D(l+k)-\Delta(l+k)\big]\right. \nonumber \\
&&\left. \hspace{-1.5cm}
+\int\!\! d^{4} l\,B(0,l^2,l^2)D(l)\Delta(l+k) 
+\frac{1}{k^2}\int\!\! d^{4} l\,(k\cdot l)B(0,l^2,l^2)D(l)D(l+k) \right\} \,.
\label{vert1}
\eea

It is convenient to express the measure in spherical coordinates,
\be
\int\!\! d^{4} l = 
2 \pi\!\!\int_{0}^{\pi} \!\!\! d\chi\sin^2\chi\, 
\int_{0}^{\infty}\!\!\! dy y \, ;
\label{spher}
\ee and rewrite (\ref{vert1}) in terms of the new variables
$x\equiv  k^2$,  $y\equiv  l^2$,  and \mbox{$z  \equiv
(l+k)^2$}.  In order to convert Eq.(\ref{vert1})
into  a one-dimensional  integral  equation,   we  resort   to  the 
standard angular approximation, defined as
\be
\int_{0}^{\pi}\!\!\! d\chi\sin^2\chi \, f(z) \approx \frac{\pi}{2} 
\bigg[\theta(x-y)f(x) + \theta(y-x)f(y)\bigg],
\label{angle}
\ee
where $\theta(x)$ is the Heaviside step function. 

Then, introducing the above change of variables and using Eq.(\ref{spher}) and (\ref{angle}) in
(\ref{vert1}), we arrive at the following linear and homogeneous equation
\bea
B(0,x,x)&=&\frac{C_{\rm A}g^2}{128\pi^2}\left\{\frac{1}{x}
\big[D(x)-\Delta(x)\big]\int_{0}^{x}\!\!\!dy\,y^2 B(0,y,y)D(y)\right.
\nonumber\\
&&\left.+\int_{x}^{\infty}\!\!\!dy\,(x-2y) B(0,y,y)D(y)\big[D(y)-\Delta(y)\big] 
+2\int_{x}^{\infty}\!\!\!dy\,y B(0,y,y)D(y)\Delta(y)
 \right. \nonumber \\
&&\left.- \frac{2}{x}D(x)\int_{0}^{x}\!\!\!dy\,y^2 B(0,y,y)D(y)
+4\,\Delta(x)\int_{0}^{x}\!\!\!dy\,y B(0,y,y)D(y) \right\}\,.
\label{vert2}
\eea

Due to the linear nature of (\ref{vert2}) it is evident that 
if $B$ is one solution then the entire family of functions $cB$, generated by 
multiplying $B$ by an arbitrary constant $c$, are also solutions.

Before embarking into the numerical treatment of (\ref{vert2}), it 
is useful to study the asymptotic solution that this 
equation furnishes for $x\to \infty$. 
In this limit on can  safely replace the  various propagators 
appearing  on the r.h.s    of     (\ref{vert2})    by    their     tree-level    values,
i.e.  $\Delta(t)\rightarrow   1/t$  and  $D(t)\rightarrow   1/t$  with
$(t=x,y)$. Then, the  first and  second terms vanish, and the leading
contribution comes from the  third term of (\ref{vert2}). Specifically, 
the asymptotic behavior of $B(0,x,x)$ is determined from the integral equation
\be
B(0,x,x)= \lambda \int_{x}^{\infty}\!\!\!dy\,\frac{B(0,y,y)}{y} \,,
\label{asym}
\ee
where $\lambda=C_{\rm A}g^2/64\pi^2$.
Eq.(\ref{asym}) can be solved easily by converting it into a first-order differential equation, 
which leads to the following asymptotic behavior
\be
B(0,x,x)= \sigma x^{-\lambda}\,,
\label{asymp_sol}
\ee
with   $\sigma$   is    an   arbitrary   parameter,   with   dimension
$[M^2]^{\lambda}$, where  $M$ is an arbitrary mass-scale.   As we will
see  in  what follows,  $\sigma$  will  be  treated as  an  adjustable
parameter, whose  dimensionality will be eventually  saturated by that
of the effective  gluon mass, or, equivalently, by  the QCD mass scale
$\Lambda$.

With the  asymptotic behavior (\ref{asymp_sol}) at hand,  we can solve
numerically the integral equation given in (\ref{vert2}). To do so, we
start  by specifying the  expressions we  will use  for the  gluon and
ghost   propagators.

As has been advocated in a series of studies based on a variety of approaches,
the gluon propagator reaches a finite value in the deep IR~\cite{Dudal:2007ch, Aguilar:2002tc}.   
This type of behavior has been observed in  Landau gauge 
in previous lattice studies~\cite{Alexandrou:2000ja},
and more recently in new, large-volume simulations~\cite{Cucchieri:2007md}.
Within the gauge-invariant truncation scheme implemented by the PT, 
the gluon propagator (effectively in the background Feynman gauge) was shown to 
saturate in the deep IR \cite{Aguilar:2006gr,Aguilar:2007ie}.  
The numerical solutions may be fitted very accurately 
by a propagator of the form
\be
\Delta(k^2) = \frac{1}{k^2 + m^2(k^2)}\,,
\label{fprog1}
\ee
where $m^2(k^2)$ acts as an effective gluon mass, 
presenting a non-trivial dependence
on the momentum $k^2$. Specifically, the mass displays 
either a logarithmic running
\be
m^2(k^2)=m^2_0\Bigg[\ln
\left(\frac{k^2+\rho\,m^2_0}{\Lambda^2}\right)\Big/\ln\left(\frac{\rho\,m^2_0}{\Lambda^2}\right) \Bigg]^{-1-\gamma_1} \,,
\label{dmass_log}
\ee 
where $\gamma_1 >0$ is the anomalous dimension of the effective mass, 
or power-law running of the form 
\be
m^2(k^2)=\frac{m^4_0}{k^2+m^2_0}\Bigg[\ln
\left(\frac{k^2+\rho\,m^2_0}{\Lambda^2}\right)\Big/\ln\left(\frac{\rho\,m^2_0}{\Lambda^2}\right) \Bigg]^{\gamma_2-1} \,,
\label{mass_power}
\ee with $\gamma_2>1$.  Which of these two behaviors  will be realized
is a delicate dynamical problem, and depends, among other things,
on the specific form of the full three-gluon vertex employed in 
the SDE for the gluon propagator
(for a detailed  discussion see \cite{Aguilar:2007ie}).  
Here we  will employ  both  functional forms,
and study the numerical impact they may have 
on the solutions of (\ref{vert2}). A plethora of phenomenological studies favor values of $m_0$ in the 
range of \mbox{$0.5-0.7\, \mbox{GeV}$}.

In addition, when solving (\ref{vert2})  an  appropriate  Ansatz   for  the  ghost
propagator $D(k^2)$  must also be furnished,  given that we  are in no
position to solve the ghost  SDE of (\ref{g1}) for arbitrary values of
the  momentum, since  this would  require  the solution  of a  coupled
system of several integral equations  involving $D$, $A$, and $B$, for
arbitrary values of the four-momenta.  Given that our aim is to study
the self-consistent  realization of an  IR finite ghost-propagator,
it is natural to employ  an Ansatz in close analogy to (\ref{fprog1}),
namely
\be
D(k^2) = \frac{1}{k^2 + M^2(k^2)} \,,
\label{ggp}
\ee
where $M^2(k^2)$ stands for a dynamically generated, effective ``ghost
mass''.  Evidently, $D^{-1}(0) = M^2(0)$, and \mbox{$D^{-1}(0)\neq 0$}
provided   that  \mbox{$M^2(0)   \neq  0$}.    Of  course,   once  the
corresponding  solutions   for  $B(0,x,x)$  have   been  obtained  the
self-consistency of  the Ansatz for $M^2(k^2)$ must  be verified.  The
way  this  will  be done  in  the  next  section is  by  substituting
$B(0,x,x)$    into    the    (properly    regularized)    
integral on the 
r.h.s.    of
Eq.~(\ref{tadghost}),  and then demanding  that  
its value is equal  to  the  $M^2(0)$ appearing on the l.h.s.

For the actual momentum dependence of the 
effective ghost  mass, $M(k^2)$ we will assume 
three  different characteristic behaviors and  will analyze the
sensitivity of $B(0,x,x)$ on them.

We will employ the following three types of $M(k^2)$:    

(i)   ``hard    mass'',    i.e.   a    constant   mass with no running,
\be
M^2(k^2)=M^2_0\,,
\label{dmass_const}
\ee

(ii)  logarithmic  running of the form
 
\be
M^2(k^2)=M^2_0\Bigg[\ln
\left(\frac{k^2+\rho\,M^2_0}{\Lambda^2}\right)\Big/\ln\left(\frac{\rho\,M^2_0}{\Lambda^2}\right) \Bigg]^{-1-\kappa_1} \,,
\label{dmass_log1}
\ee 

(iii)  power-law   running, given by 
\be
M^2(k^2)=\frac{M^4_0}{k^2+M^2_0}\Bigg[\ln
\left(\frac{k^2+\rho\,M^2_0}{\Lambda^2}\right)\Big/\ln\left(\frac{\rho\,M^2_0}{\Lambda^2}\right) \Bigg]^{\kappa_2-1} \,.
\label{dmass_power}
\ee 
Clearly, the last two possibilities, 
(\ref{dmass_log1}) and (\ref{dmass_power}),
 are exactly analogous to the corresponding two types of 
running of the gluon mass, (\ref{dmass_log})    
and (\ref{mass_power}), respectively.

We then solve numerically Eq.(\ref{vert2}) using the gluon and ghost
propagators    given    by    Eqs.(\ref{fprog1})   and    (\ref{ggp}),
respectively, supplemented by the various types of running 
for $m^2(k^2)$ and $M^2(k^2)$.
The  integration range is  split in two  regions, $[0,s]$
and  $(s,\infty]$,  where  $s\gg   \Lambda^2$.  For the second interval
we  impose the asymptotic  behavior of (\ref{asymp_sol}), 
choosing a value for $\sigma$.

\begin{center}
\begin{figure}[ht]
\includegraphics[scale=1.0]{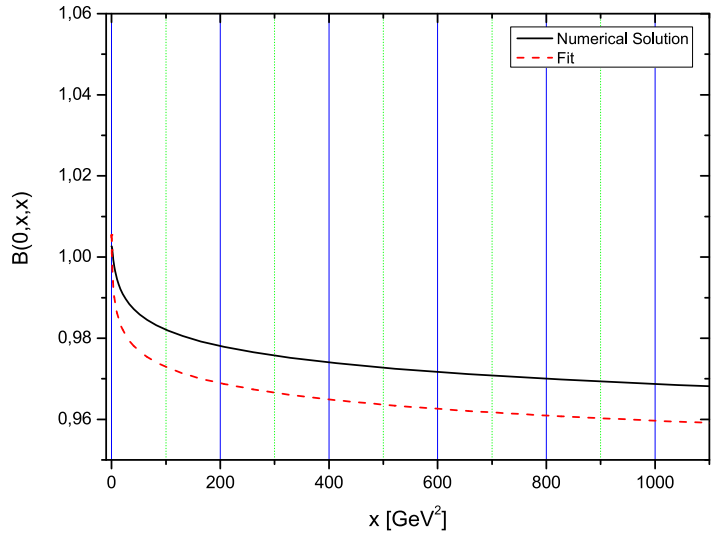}
\caption{The   black solid line  is the  numerical   solution  of
 Eq.(\ref{vert2}), assuming logarithmic type of running 
for $m^2(k^2)$ and $M^2(k^2)$, with  
$\gamma_1 = \kappa_1=0.6$, $m^2_0=0.35 \,\mbox{GeV}^2$,  
$M^2_0=0.4 \,\mbox{GeV}^2$, 
$\rho=4$, and  $\sigma^{1/\lambda}= 1 \mbox{GeV}^{2}$.
The red dashed line represents the 
 fit of Eq.(\ref{fit_b}); the relative difference between the two
curves is  less than $1\%$ 
(note the fine spacing of the $y$  axis).}
\label{f6}
\end{figure}
\end{center}

It turns  out that the  numerical solution obtained for  $B(0,x,x)$ is
rather insensitive to the form of the gluon mass employed, and it mainly
depends on  the form of the ghost  propagator. 
More specifically, we  can fit
the numerical solution with an impressive accuracy 
by means of the simple, physically motivated function
\be
B(0,x,x)= \frac{\sigma}{[x+M^2(x)]^{\lambda}} \,,
\label{fit_b}
\ee
regardless of the form of momentum dependence employed for  $M^2(x)$.   
Evidently, for large  values of $x$ the  above expression
goes over the asymptotic solution of Eq.(\ref{asymp_sol}).
In Figs.(\ref{f6}),  we present a typical solution  for $B(0,x,x)$ 
together with the fit given by (\ref{fit_b}).

%

\section{Infrared finite ghost propagator}
\label{Sect:comb}

In the previous section we have 
obtained the general 
solutions for $B(0,x,x)$, under the assumption that the 
ghost propagator was finite in the IR, and more 
specifically that it was given by the general form of (\ref{ggp}). 
The next crucial step consists in substituting the solutions 
obtained for $B(0,x,x)$ into (\ref{tadghost}) and in 
examining under what conditions the two hand sides of the equation 
can be made to be equal. As we will see this procedure will 
eventually boil down to constraints on the values that one is allowed  
to choose for the free parameter $\sigma$.

Substituting Eqs.(\ref{ggp}) and (\ref{fit_b}) into (\ref{tadghost}), we arrive at
\be
D^{-1}(0) =-C_{\rm A}g^2\sigma\int[dk]\frac{1}{[k^2+M^2(k^2)]^{1+\lambda}}\,.
\label{td1}
\ee
The r.h.s. of (\ref{td1}) is simply given by 
\be
D^{-1}(0) = M^2_0 \,,
\ee
for any form of $M^2(k^2)$. 
Let us first verify the self-consistency of (\ref{td1}) for the 
case where the ghost mass vanishes identically, i.e. $M^2(k^2)=0$.
Then, (\ref{td1}) reduces to nothing but the   
standard dimensional regularization result~\cite{Collins}
\be
\int[dk](k^2)^{-\alpha}=0\,,
\label{dreg}
\ee
valid for any value of $\alpha$, for the special value $\alpha = 1+\lambda$.

For non-vanishing $M^2(k^2)$ the integral on the r.h.s. of (\ref{td1}) 
is UV divergent:
at large $k^2$  it goes as $(\Lambda_{{\rm UV}})^{1+\lambda}$, 
where $\Lambda_{{\rm UV}}$ is a UV momentum cutoff. 
It turns out that the r.h.s. can be made UV finite by simply subtracting
from it its perturbative value, i.e. the vanishing integral of (\ref{dreg})
~\cite{foot2}.

Carrying out this regularization procedure explicitly, one obtains 
\bea
M^2_0 &=& - C_{\rm A}g^2\, \sigma \int[dk]\left(\frac{1}{[k^2+M^2(k^2)]^{1+\lambda}}
 - \frac{1}{(k^2)^{1+\lambda}} \right) \nonumber \\ 
&=& -C_{\rm A}g^2\sigma\int\frac{[dk]}{[k^2+M^2(k^2)]^{1+\lambda}}\left(1-\left[1+\frac{M^2(k^2)}{k^2}\right]
^{1+\lambda}\right) \,.
\label{dreg1}
\eea

It is now elementary to verify that the integral on the r.h.s of (\ref{dreg1}) converges.
At large $k^2$
we can expand the second term in the parenthesis and neglecting in the denominator $M^2(k^2)$ next to $k^2$,
we find that the resulting integral (apart of multiplicative factors) is given by
\be
\int \!\!dy\frac{M^2(y)}{y^{1+\lambda}}\,.
\label{iuv}
\ee
Notice that the above integral converges even for the less favorable 
case of a constant $M^2(y)$; then, (\ref{iuv})  is  proportional  to $y^{-\lambda}$,  
and is therefore convergent, since $\lambda >0$. Clearly, when 
 $M^2(y)$ drops off in the UV, as described by (\ref{dmass_log1}) or (\ref{dmass_power}),
the integral converges even faster.
Next we will analyze separately what happens for
each one of the three different Ans\"atze 
we have employed for $M^2(y)$, Eqs.(\ref{dmass_const}) -- (\ref{dmass_power}).

The case of a constant ghost mass can be easily worked out. 
Replacing \mbox{$M^2(k^2)\rightarrow M_{0}^2$} in Eq.(\ref{dreg1}),  
keeping only the leading contribution to the integral, we arrive at
(notice the cancellation of the  coupling constant $g^2$ 
appearing in  front of the integral)
\be
M^2_0= \frac{4 \sigma}{1-\lambda} M_0^{2(1-\lambda)}  \,.
\label{hard_tad}
\ee
Then, in order to enforce the equality of both sides  of (\ref{hard_tad}) 
$\sigma$ must satisfy 
\be
\sigma=\frac{(1-\lambda)}{4} M_0^{2\lambda} \,.
\label{cdep}
\ee
Evidently,  $\sigma$ depends very weakly on  $M_0$, 
and its value is practically fixed at $1/4$. 
Indeed, given that $\lambda$  is a small number,  of the order of ${\mathcal O}(10^{-2})$, 
Eq.(\ref{cdep}) may be expanded as  
\be
\sigma \approx \frac{(1-\lambda)}{4}\, \Lambda^{2\lambda} \,\left[1+\lambda\ln\left(\frac{M_0^2}{\Lambda^2}\right)\right]\,,
\label{exp_cdep}
\ee
%
\begin{center}
\begin{figure}[ht]
\includegraphics[scale=1.0]{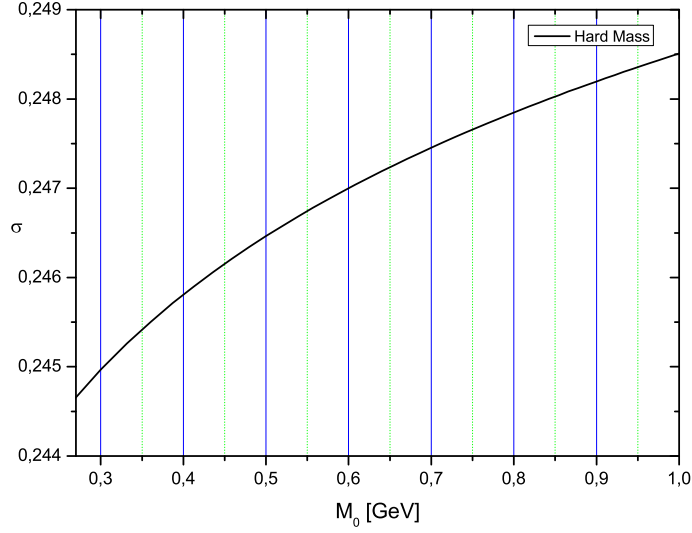}
\caption{$\sigma$ as a function of the hard ghost mass $M_0$, obtained from  Eq.(\ref{cdep}).} 
\label{f1}
\end{figure}
\end{center}
from where it is clear that $\sigma$ can only assume values slightly different of $1/4$.
In Fig.(\ref{f1}), we show this mild dependence
of $\sigma$ on $M_0$, for $\Lambda=300 \,\mbox{MeV}$.

\begin{center}
\begin{figure}[hb]
\includegraphics[scale=1.0]{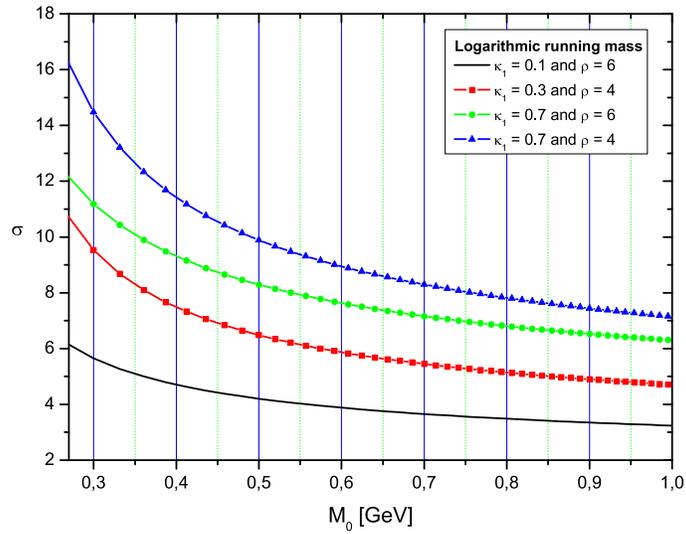}
\caption{$\sigma$ as function of $M_0$, when $M^2(k^2)$ runs   
logarithmically, as in Eq.(\ref{dmass_log1}).}
\label{f2}
\end{figure}
\end{center}

\begin{center}
\begin{figure}[ht]
\includegraphics[scale=1.0]{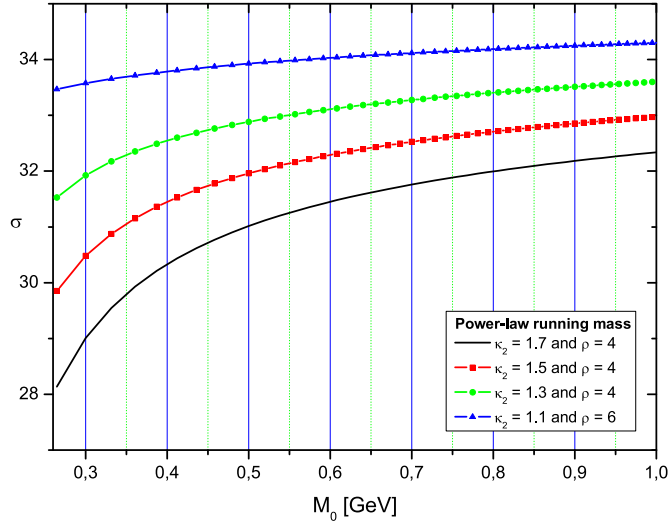}
\caption{$\sigma$ as function of $M_0$,  when the 
 power-law running of Eq.(\ref{dmass_power}) is assumed for  $M^2(k^2)$.}
\label{f3}
\end{figure}
\end{center}

We next turn to the case where $M^2(y)$ displays the logarithmic or power-law dependence 
on the momentum, described by Eqs.~(\ref{dmass_log1}) and (\ref{dmass_power}), respectively.
Now the integrals cannot be carried out analytically and have been computed  numerically.
Choosing different values for  $\kappa_1$, $\kappa_2$, and $\rho$, 
we obtain the curves presented in Fig.(\ref{f2}) and Fig.(\ref{f3}),
showing the dependence of $\sigma$ on $M_0$.
%

Several observations are in order:

(i) For both types of running the results show a   
stronger dependence on $M_0$ than in the case of the hard mass. 

(ii) The range of possible values for $\sigma$ increases  
significantly. Whereas in the case of constant mass 
one was practically restricted to a unique 
value for $\sigma$, namely $\sigma \approx 1/4$ [viz. Fig.(\ref{f1})], 
now one may obtain self-consistent solutions 
choosing values for $\sigma$ over a much wider interval.

\begin{center}
\begin{figure}[ht]
\includegraphics[scale=1.0]{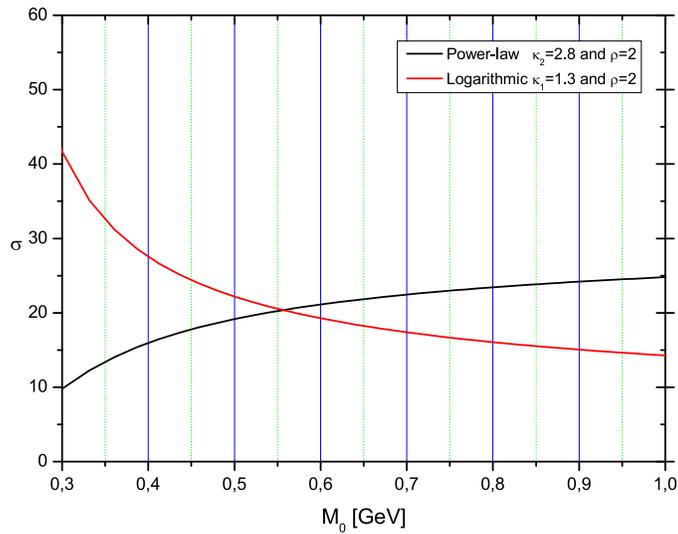}
\caption{For $\sigma=20$, a ghost mass 
can be generated from Eq.(\ref{dreg1}) for either type of running.}
\label{f4}
\end{figure}
\end{center}

(iii) There is a qualitative difference between the logarithmic and power-law 
running: in the former case $\sigma$ is a decreasing function of $M_0$, 
while in the latter it is increasing.
This 
offers the particularly interesting possibility of finding values for 
$\sigma$ that  furnish self-consistent solutions for either types of 
running of $M^2(k^2)$. 
A characteristic example where Eq.(\ref{dreg1}) is satisfied for the same 
value of $M_0$ for both types of running 
is shown in Fig.(\ref{f4}): for  $\sigma\approx 20$ one may 
generate
a ghost mass of $M_0\approx 560 \,\mbox{MeV}$,  assuming 
for $M^2(k^2)$ either the logarithmic running  of 
Eq.(\ref{dmass_log1}), or the power-law running of Eq.(\ref{dmass_power}).

%
%
%

\setcounter{equation}{0}
\section{Discussion and Conclusions}
\label{Sect:Concl}

In this  article we  have demonstrated that  it is possible  to obtain
from  the SDEs  of QCD  an IR  finite ghost  propagator in  the Feynman
gauge.   In  this  construction  the  longitudinal  component  of  the
gluon-ghost vertex, which  is inert  in the  Landau gauge,  assumes a
central  role,  allowing  for  $D(0)$  to  be  finite.   This  is
accomplished without  having to assume  any special properties  of the
form-factor, other than a nonvanishing limit in the IR; in particular,
we do  not need to impose the  presence of massless poles  of the type
$1/p^2$.

Our procedure may be summarized as follows.  First of all, since we are
interested in  the possibility of obtaining $D^{-1}(0)\neq  0$ we have
focused  on  the form  of  the  ghost gap  equation  in  the limit  of
vanishing external  momentum, $p\to 0$.   Next we have  linearized the
SDE  for  the  form-factor  $B(p^2,q^2,k^2)$  , and  have  looked  for
solutions for  the special kinematic configuration  of vanishing ghost
momentum,  $B(0,k^2,k^2)$,  which  is   relevant  for  the  ghost  gap
equation. The  solution may be fitted  in the entire  range of momenta
with a particularly simple, physically motivated expression.  Coupling
the two equations together,  we have obtained the conditions necessary
for self-consistency.  It essentially  boils down to relations between
the  free  parameter  $\sigma$  and  the  values  of  $D^{-1}(0)$,  or
equivalently  $M_0^2$,  as  captured  in  Figs~(\ref{f1})--(\ref{f3}).
These figures  furnish the  value of $M_0$  one obtains if  a concrete
value of $\sigma$ is chosen, assuming certain characteristic types of
running for the the ghost mass function $M^2(k^2)$.
The freedom in  choosing the value of $\sigma$ will  be restricted, or
completely  eliminated,  in  the  non-linear  version  of  the  vertex
equation. It  would certainly  be interesting to  venture into  such a
study,  because it  is  liable to  pin  down completely  the value  of
$D^{-1}(0)$.

The most immediate physical  implication of the results presented here
is  that the  finite gluon  propagator  obtained in  the previous  SDE
studies  in  the  PT-BFM   framework,  with  the  ghost  contributions
gauge-invariantly omitted, will not  get destabilized by the inclusion
of the ghost loops.  Specifically,  one would expect that the addition
of the  ghost loop  into the corresponding  SDE should not  change the
qualitative picture.  The quantitative  changes induced should also be
small; mainly the correct coefficient of $11 C_A/48 \pi^2$ multiplying
the  renormalization group  logarithms will  be restored  (without the
ghosts  it is  $10 C_A/48  \pi^2$), and  it might  inflate  or deflate
slightly the  corresponding solutions for the gluon  propagator in the
intermediate  region  between $0.1-1\,\mbox{GeV}^2$.  Of  course, 
a complete
analysis  of  the coupled  SDE  system is  needed  in  order to  fully
corroborate this general picture.

Given the  complexity and importance of  the problem at  hand it would
certainly  be essential to confront  these SDE  results  with lattice
simulations  of  the  ghost  propagator  in  the  Feynman  gauge.   In
addition, since  the formulation  of the BFM  on the lattice  has been
presented long  ago by  Dashen and Gross 
(in the  Feynman gauge)~\cite{Dashen:1980vm}, and has already been 
used~\cite{DiGiacomo:1987tf, Trottier:1997bn},  
one might also consider the possibility of simulating the gluon propagator
within  that particular  gauge-fixing scheme,  thus enabling  a direct
comparison with the SDE results predicting an IR finite answer.

\section*{Acknowledgments}

This work was supported by the Spanish MEC under the grants FPA 2005-01678 and 
FPA 2005-00711.  
The research of JP is funded by the Fundaci\'on General of the UV. 


\begin{thebibliography}{99}
\bibitem{foot2}
This is a rather standard operation in dimensional regularization.  
Its most familiar version is nothing but the usual statement that 
in dimensional regularization there are no quadratic divergences,
simply because  $\int[dk]/k^2 =0$,
and therefore 
\be
\int\frac{[dk]}{k^2+m^2} - \int\frac{[dk]}{k^2} = - m^2 \int \frac{[dk]}{k^2(k^2+m^2)}
\nonumber
\ee
\end{thebibliography}

\end{document}